\documentclass{article}
\usepackage[T1]{fontenc} 
\usepackage[utf8]{inputenc} 
\usepackage{ismir,amsmath,cite,url,amsfonts}
\usepackage{graphicx}
\usepackage{color}
\usepackage{booktabs}
\usepackage{url}
\usepackage{mathtools}
\usepackage{etoolbox,siunitx}
\usepackage{multirow}
\usepackage[table,xcdraw]{xcolor}
\usepackage{pifont}

 \renewcommand{\bfseries}{\fontseries{b}\selectfont}
\newrobustcmd{\B}{\bfseries}

\newcommand{\setmeter}[2]{\ensuremath{%
  \vcenter{\offinterlineskip
    \halign{\hfil##\hfil\cr
            $\scriptstyle#1$\cr
            \noalign{\vskip1pt}
            $\scriptstyle#2$\cr}
  }}%
}

\title{The Chamber Ensemble Generator: \\ Limitless High-Quality MIR Data via Generative Modeling}

\multauthor
{Yusong Wu$^1$ \hspace{1cm} Josh Gardner$^{2, 3}$ \hspace{1cm} Ethan Manilow$^3$} { \bfseries{Ian Simon$^3$ \hspace{1cm} Curtis Hawthorne$^3$ \hspace{1cm} Jesse Engel$^3$} \\
 $^1$ Université de Montréal, Mila\\
$^2$ University of Washington\\
$^3$  Google Research, Brain Team\\
{\tt\small }
}

\def\authorname{Y. Wu, J. Gardner, E. Manilow, I. Simon, C. Hawthorne, J. Engel}

\usepackage[bookmarks=false,pdfauthor={\authorname},pdfsubject={\papersubject},hidelinks]{hyperref}

\begin{document}

\maketitle
\begin{abstract}
Data is the lifeblood of modern machine learning systems, including for those in Music Information Retrieval (MIR). However, MIR has long been mired by small datasets and unreliable labels. In this work, we propose to break this bottleneck using generative modeling. By pipelining a generative model of notes (Coconet trained on Bach Chorales) with a structured synthesis model of chamber ensembles (MIDI-DDSP trained on URMP), we demonstrate a system capable of producing unlimited amounts of realistic chorale music with rich annotations including mixes, stems, MIDI, note-level performance attributes (staccato, vibrato, etc.), and even fine-grained synthesis parameters (pitch, amplitude, etc.). We call this system the \textbf{Chamber Ensemble Generator (CEG)}, and use it to generate a large dataset of chorales from four different chamber ensembles (CocoChorales). We demonstrate that data generated using our approach improves state-of-the-art models for music transcription and source separation, and we release both the system and the dataset as an open-source foundation for future work in the MIR community.

\end{abstract}

\section{Introduction}\label{sec:introduction}

As deep learning systems become the go-to choice for solving more and more Music Information Retrieval (MIR) tasks, it behooves researchers to lean into the strengths of these systems. For example, it is now well-established that neural networks perform better when they are larger and have access to more data~\cite{kaplan2020scaling,brown2020language,henighan2020scaling}. However, the bottleneck to scaling MIR systems is a lack of data with high-quality labels. If MIR researchers could access larger quantities of labeled data, we could scale our systems to further increase their ability to understand and generate musical audio.

In fact, MIR tasks that \textit{do} have abundant data--such as music tagging~\cite{bogdanov2019mtg,won2019self-attn-tagging,won2020evaluation} or piano transcription~\cite{hawthorne2018enabling,kong2021high,hawthorne2021sequence}--have already seen large performance gains with architectures (e.g., Transformers~\cite{vaswani2017attention}) that can take advantage of large-scale data.
For other tasks, researchers have found clever ways to scale MIR systems by using unsupervised generative modeling~\cite{huang2018music,payne2019musenet,dhariwal2020jukebox} thus loosening requirements for labeled data and enabling training on larger corpora.
Researchers have also found ways to leverage these large unsupervised models for downstream MIR tasks~\cite{castellon2021codified,manilow2022source}.

A common way to combat data scarcity is data augmentation. Data augmentation is a necessity for some MIR tasks like source separation~\cite{schluter2015exploring,uhlich2017improving,manilow2019cutting,defossez2019music,pretet2019singing,manilow2020open,kong2021decoupling,song2021catnet}, and is highly effective for others like self-supervised learning for classification ~\cite{spijkervet2021contrastive,won2021semi,wang2022towards} and pitch estimation~\cite{gfeller2020spice}. However, augmentation relies on the labels provided with the original data; if those labels are small or unreliable, augmentation may not be helpful. Furthermore, even with perfect labels there is a limit to how much we can scale using augmentation alone. The amount and types of perturbations we can apply to musical data cannot be done indiscriminately because augmentations must preserve the semantic link between the audio content and labels.



One approach to realize the promise of large models is \textit{dataset amplification},\footnote{By ``amplification'' we do not mean using an electronic amplifier (or amp simulation software), but rather the process of \emph{expanding} a dataset using generative models.} whereby generative models are used to create larger datasets from existing data, thus ``amplifying'' small amounts of data~\cite{jahanian2022generative, zhang2021datasetgan, christiano2018supervising, axelrod2020sample}.
Amplifying data is particularly auspicious in MIR tasks, where data is often costly to collect and label. Given a sufficiently good generative model trained on a small dataset, one could use the model to create a large amount of data with paired annotations. This data can then be used to train other large models, which would be impossible with the original dataset.

\begin{figure*}[!t]
 \centerline{
 \includegraphics[width=\textwidth]{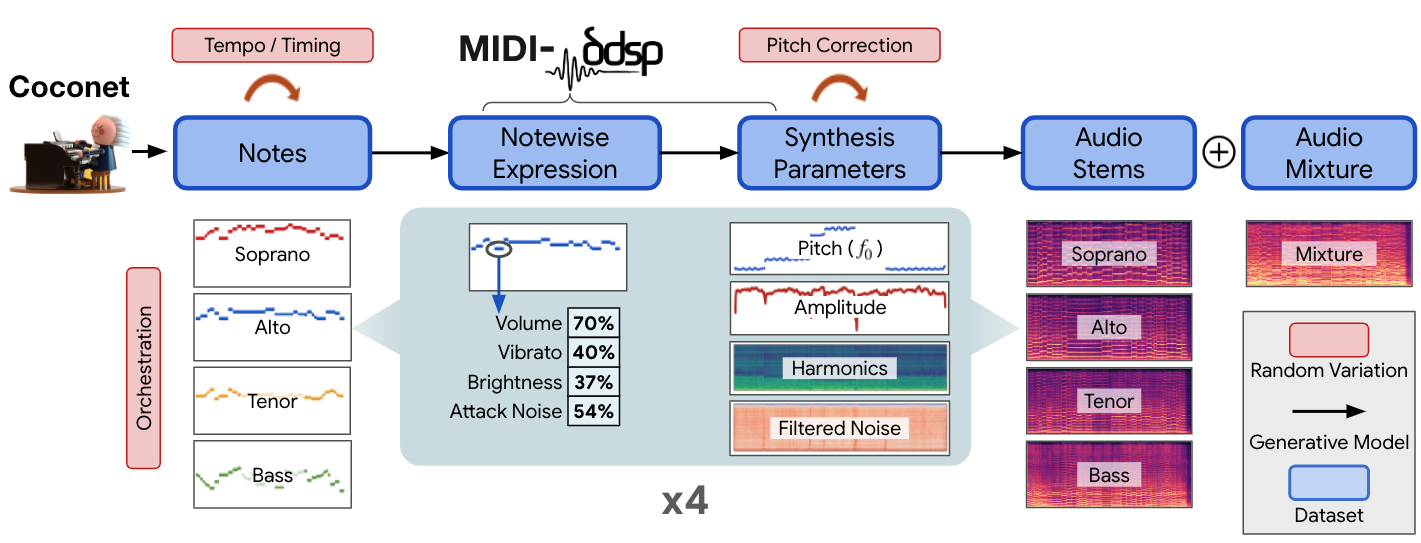}}
 \caption{The \textbf{Chamber Ensemble Generator (CEG)} generates a dataset of realistic-sounding performances of Bach Chorales by pipelining two generative models, CocoNet~\cite{huang2019counterpoint} and MIDI-DDSP~\cite{wu2022mididdsp}. Because these models form a structured hierarchy, we are able to produce data at many different stages in the generation process, including aligned note data with instrument labels, notewise expression data, synthesis parameter data, audio stem data, and audio mixture data. The structured hierarchy also enables us to manually add manipulation at certain stages to introduce variation. In the dataset accompanying this paper, \textbf{CocoChorales}, we vary the orchestration of the pieces and overall tempo, we alter the microtiming of notes, and we randomly apply pitch correction to the performances.} 
 \label{fig:hero-diagram}
\end{figure*}

In this paper, we present the \textbf{Chamber Ensemble Generator (CEG)}, a pipeline of generative models which we use to create a large synthetic audio dataset. The CEG is a dataset amplification system built on generative models that are trained on small datasets (426 total examples). Specifically, we use a set of \textit{structured} generative models, i.e., models that have interpretable intermediate representations. Using structured models produces many types of high-quality labels and allows us to manipulate the generative processes in many different ways. Specifically, we use a music composition model, Coconet~\cite{huang2019counterpoint}, to generate four-part note data in the style of Bach Chorales, and we use an audio generation model, MIDI-DDSP~\cite{wu2022mididdsp}, to turn the note data into audio for each instrument in the ensemble. We used the CEG to generate $240,000$ performances containing audio mixture data with high-quality annotations for stems, MIDI, note-level performance attributes, and fine-grained synthesis parameters. We call the resulting dataset \textbf{CocoChorales}, and we make both data and code publicly available.\footnote{\url{https://g.co/magenta/ceg-and-cocochorales/}} We show that state-of-the-art models for music transcription and source separation benefit greatly from the expanded dataset and accompanying labels.

\section{Related Work}


Training on synthetic data to help real-world downstream applications has long been used in Machine Learning, and Reinforcement Learning \cite{engel2017neural, nikolenko2021synthetic, zhao2020sim, tremblay2018falling, axelrod2020sample}.
Recently, the use of generative models for dataset amplification, i.e., enlarging a dataset by generating large synthetic datasets, has become more common in vision and other domains \cite{jahanian2022generative, zhang2021datasetgan, christiano2018supervising, wood2021fake}.
In the image domain, Zhang et al. \cite{zhang2021datasetgan} trained an additional semantic map decoder on the latent space of a pre-trained StyleGAN \cite{karras2019style}. By training the decoder on a few annotated examples, the model becomes an ``infinite dataset generator'' with paired semantic maps, achieving the similar performance as a model trained on 100x more human-annotated data. Here, we implement a similar approach in the music domain, with the hope of enlarging the effective size of our training data. Specifically, we train a set of generative models from scratch instead of using a pre-trained model.

Synthetic datasets have been proposed a handful of times in the music domain.
The JS Fake Chorales dataset \cite{peracha2021js} consists of 500 synthetic Bach Chorales generated by an RNN-based generative model. However, the JS Fake Chorales dataset is small compared to this work and consists of symbolic note data, lacking any audio performance data. Similarly, Liu et. al.~\cite{liu2020incorporating} propose a system to grade model's generated data and feed high-quality data from the model back into its training set. This work, also, only used symbolic data and no audio performance data. The Synthesized Lakh (Slakh) Dataset \cite{manilow2019cutting} is constructed by synthesizing 2100 MIDI files using professional-grade sample-based synthesizers, resulting in 145 hours of mixture audio with paired stem audio and MIDI files.
Other work has also proposed using sample-based synthesizers for source separation specifically~\cite{miron2017generating,manilow2020bespoke,chen2022improving} or drum transcription~\cite{cartwright2018increasing}.
However, the output of these synthesizers can lack finer-grained annotations like $f_0$'s for each note, and--most importantly--due to the limitations of automating these synthesizers for large scale data creation, the resulting audio does not sound like it was performed by live musicians. In other words, there is a risk of a mismatch between the distributions of the synthesized data and real world data. Here, we use generative models to produce highly realistic audio as a step towards mitigating this potential distribution mismatch.

\begin{table*}[!ht]
\centering
\begin{tabular}{@{}lcrrc@{}}
\toprule
\multicolumn{1}{c}{Name} & Instrumentation  & \# Examples   & {Duration (hrs)}   & Content \\  \midrule 
\multicolumn{5}{c}{\cellcolor[HTML]{EFEFEF}\textit{\textbf{MIDI-only Datasets}}} \\
Bach Chorales~\cite{boulanger2012modeling}  & 4-part chorale & 382    & 26  & MIDI \\
JS Fake Chorales~\cite{peracha2021js}  & 4-part chorale & 500       & 1     & MIDI  \\

Lakh MIDI~\cite{raffel2016learning}   & 128 MIDI instruments & 176,581    & 10,521    & MIDI \\
Meta MIDI~\cite{ens2021building}   & 128 MIDI instruments & 436,631  & 19,224    & MIDI  \\ [0.5em]
\multicolumn{5}{c}{\cellcolor[HTML]{EFEFEF}\textit{\textbf{Audio \& MIDI Datasets -- Single Instrument}}} \\
MAPS~\cite{emiya2010maps}  & Piano  & 270  & 18  & Audio, MIDI \\
MAESTRO~\cite{hawthorne2018enabling}    & Piano    & 1,276    & 199 & Audio, MIDI \\
GuitarSet~\cite{xi2018guitarset}   & Guitar   &  360  &  3 & Audio, MIDI, $f_0$'s, Tempo, Chords \\
FiloSax~\cite{foster2021filosax}     & Saxophone     & 240     & 24       & Audio, MIDI  \\ [0.5em]
\multicolumn{5}{c}{\cellcolor[HTML]{EFEFEF}\textit{\textbf{Audio \& MIDI Datasets -- Ensembles}}} \\
MUSDB18~\cite{musdb18} & 4 instruments  & 150  & 10  & Mix and stem audio only \\
MusicNet~\cite{thickstun2016learning}  & 11 instruments & 330    & 34    & Mix audio, MIDI \\
URMP~\cite{li2018creating}  & 14 instruments & 44  & 1 & Mix and stem audio, video, MIDI  \\
Slakh~\cite{manilow2019cutting}  & 34 instruments  & 2,100  & 145 & Mix and stem audio, MIDI \\ [0.5em]
\multicolumn{5}{c}{\textit{CocoChorales (this work)}}  \\ \midrule
String    & String ensemble    & 60,000  & 350  &  \\
Brass     & Brass ensemble     & 60,000  & 350  &  \\
Woodwind  & Woodwind ensemble  & 60,000  & 350  &  \\
Random    & Random instruments & 60,000  & 350  &  \\ [0.25em]
\quad Total     & 13 instruments       & 240,000              & 1,400               & \multirow{-5}{*}{\begin{tabular}[c]{@{}l@{}}\phantom{ + }Mix and stem audio, MIDI \\ + Note: volume, vibrato, attack, ... \\ + Synthesis: $f_0$'s, loudness, noise, ...\end{tabular}} \\ \bottomrule
\end{tabular}
\caption{This table compares our work, CocoChorales (last 5 rows), to a selection of existing datasets. Durations are counted in hours of mixture data (where applicable) and are rounded to the nearest hour. }
\label{tab:comparison}
\end{table*}

There are a large number of terrific options for generative music models that could be used to amplify or generate a dataset.
For instance, the output quality of general purpose audio generation models has steadily increased in recent years~\cite{oord2016wavenet, kalchbrenner2018efficient, engel2019gansynth, dhariwal2020jukebox, goel2022s, borsos2022audiolm}, however, controlling fine-grained aspects of the performance (e.g., whether vibrato is applied to an individual note) is difficult with these models. Therefore, we must look elsewhere if we desire to use such models to make datasets with detailed annotations.
This leads us toward using a set of models for structured generation, each with a designated job. For instance, one could use a composition model (e.g., a piano \cite{huang2018music}, singing \cite{ju2021telemelody, chen2020melody}, or symphonic \cite{liu2022symphony} composition model) to generate notes, and sonify its output with a score-to-audio synthesis model \cite{wang2019performancenet, jonason2020control, castellontowards, dong2022deep}. Still, few score-to-audio models enable low-level performance details, like the vibrato of a single note.
In this work we chose Coconet~\cite{huang2019counterpoint} and MIDI-DDSP~\cite{wu2022mididdsp} because of the high amount of structure these models contain (i.e., Coconet \textit{only} generates four-part Bach chorales, and MIDI-DDSP has a 3-level interpretable hierarchy). These models enable more control and variation of the generation process (e.g., ensemble types, notewise performance characteristics, etc) and, thus we are able to create a dataset with more types of labels to support more tasks.

\section{The Chamber Ensemble Generator}
 
In this section, we will first describe the two generative models used in our Chamber Ensemble Generator (CEG), which is used to create the CocoChorales dataset.
An overview of the CEG is illustrated in Figure \ref{fig:hero-diagram}.

\subsection{Coconet}
\label{subsec:coconet}

Coconet \cite{huang2019counterpoint} is a music composition model that generates four-part harmonic note data in the style of a Bach chorale. Coconet is trained on the J.S. Bach Chorales dataset \cite{boulanger2012modeling, jsb-dataset}, which consists of 382 pieces. Coconet trains a convolutional neural network to complete partial musical scores by infilling a randomly masked input score. During inference, Coconet iteratively applies blocked Gibbs sampling to rewrite its generation for a fixed number of steps before producing its final output score. For more details, we refer interested readers to the original Coconet paper~\cite{huang2019counterpoint}.

For the current work, we used an open-source implementation of Coconet \cite{coconet-pytorch}. However, our model slightly differs from the original Coconet in the following ways: $(1)$ We apply pitch augmentation during training by randomly transposing the input by $\{-3,-2,-1,0,1,2,3\}$ semitones. $(2)$ The loss used during training is simply unweighted cross-entropy, compared to the reweighted loss in the original paper~\cite{huang2019counterpoint}. $(3)$ Our model does not generate rests.

\subsection{MIDI-DDSP}\label{subsec:midi-ddsp}

MIDI-DDSP \cite{wu2022mididdsp} is a score-to-audio generation model which uses a three-level structured hierarchy (notes, performance, synthesis) when generating single-note audio.
Given input MIDI, the audio synthesis via MIDI-DDSP proceeds as follows. First, MIDI-DDSP generates a set of ``note expression'' characteristics for each note, each controlling one aspect of a note's performance: volume, volume fluctuation, volume peak position, vibrato, brightness, and attack noise. 
Second, MIDI-DDSP uses the note expressions and the accompanying MIDI to generate frame-wise synthesis parameters. The synthesis parameters of an audio clip consist of a fundamental frequency $f_0$, an amplitude curve, a distribution of amplitudes for each harmonic frequency above the $f_0$, and a set of filtered noise magnitudes. 
Finally, the Differentiable Digital Signal Processing (DDSP) \cite{engel2020ddsp} modules synthesize a waveform using the generated synthesis parameters. Figure \ref{fig:hero-diagram} provides an overview of this process, but we refer readers to the original MIDI-DDSP paper for further details~\cite{wu2022mididdsp}.

The intermediate representations (i.e., note expressions and synthesis parameters) generated by MIDI-DDSP can provide rich annotations for many MIR tasks. For example, the $f_0$ curves can be used for $f_0$ estimation, and the note expression can be used for performance analysis.

MIDI-DDSP is trained on the URMP dataset \cite{li2018creating}, which consists of 3.75 hours of solo recordings (1 hour of mixture data, as shown Table~\ref{tab:comparison}). MIDI-DDSP is thus capable of generating the 13 common orchestral instruments present in URMP: violin, viola, cello, double bass, flute, oboe, clarinet, saxophone, bassoon, trumpet, horn, trombone, and tuba.

\section{CocoChorales}
\label{sec:cocochorales}

Using our Chamber Ensemble Generator pipeline, we generated a dataset which we call \textbf{CocoChorales}. CocoChorales consists of $240,000$ pieces, totaling 1411 hours of mixture data. The CocoChorales is orders of magnitude larger than existing MIR datasets \cite{li2018creating, musdb18, foster2021filosax, boulanger2012modeling, jsb-dataset, thickstun2016learning, manilow2019cutting} (see Table~\ref{tab:comparison}). Every example in the dataset contains MIDI data, note expression data, synthesis parameter data, and audio for each instrument stem, audio of the mixture, and additional metadata about tempi, ensemble type, etc. We make train/valid/test splits of CocoChorales using $80\%$/$10\%$/$10\%$ portion of the overall data, respectively.
Further information about CocoChorales--including download links--can be found in the online supplement.\footnote{\url{https://g.co/magenta/ceg-and-cocochorales/}} The rest of this section is dedicated to describing the creation process for CocoChorales.

\subsection{MIDI Generation and Augmentation}
\label{subsec:midi}

Coconet \cite{huang2018music} is a generative model of Bach Chorales. We use Coconet to compose 8 measures of a standard four-part chorale (Soprano, Alto, Tenor, Bass, or SATB) in \setmeter{4}{4} time. We generate samples from Coconet by running 1024 sampling steps.
In order to ensure that each part generated by Coconet falls into the range of expected pitches for the given SATB part (e.g., a soprano note should not be too low), we reject a generated sample piece if any of the pitches fell 3 semitones outside of the min/max pitch used in the J.S. Bach Chorales dataset~\cite{jsb-dataset} for that part. 

Once we have a set of valid chorales as MIDI, we augment this data to add more variety to the dataset.
Because we are operating on MIDI data rather than audio data, we can apply tempo and timing variations without worrying about a time-stretching or pitch-shifting algorithm introducing artifacts. To that end, we randomly set the tempo (in BPM) to an integer drawn from $\textrm{Uniform}([50, 150])$. Furthermore, the raw output of Coconet is quantized at the granularity of 16$^{th}$ notes, so to add an extra level of expressiveness to the MIDI, we add microtiming offsets to the notes. Following the observation that human timing approximates a normal distribution \cite{sogorski2018correlated, naveda2011microtiming}, we add a random timing offset to each note sampled from a truncated normal distribution between $[-50, 50]$ ms with $\mu=0$ ms, $\sigma=15$ ms. 

The final way that we add variation to the MIDI data is by changing the orchestration of the ensembles. We define four ensembles--String, Brass, Woodwind, and Random--with $60,000$ examples in each. The first three ensembles have a fixed orchestration throughout, and the random ensemble has a varied orchestration, with instruments randomly selected from a pool of instruments according to the pitch range of the SATB part.
These ensembles and their instrumentation for Soprano, Alto, Tenor, and Bass, respectively, are defined as:

\begin{itemize}
    \item \textbf{String}: Violin 1, Violin 2, Viola, Cello.
    \item \textbf{Brass}: Trumpet, French Horn, Trombone, Tuba.
    \item \textbf{Woodwind}: Flute, Oboe, Clarinet, Bassoon.
    \item \textbf{Random}: Each SATB part is randomly assigned an instrument according to the following:
    \begin{itemize}
        \item Soprano: Violin, Flute, Trumpet, Clarinet, Oboe.
        \item Alto: Violin, Viola, Flute, Clarinet, Oboe, Saxophone, Trumpet, French Horn.
        \item Tenor: Viola, Cello, Clarinet, Saxophone, Trombone, French Horn.
        \item Bass: Cello, Double Bass, Bassoon, Tuba.
    \end{itemize}
\end{itemize}

Once a composition is generated by Coconet, a tempo is chosen, microtiming is added to the notes, and the orchestration is determined. The MIDI dataset is then given to MIDI-DDSP to synthesize into audio performances.

\subsection{Audio Synthesis and Mixing}
\label{subsec:expression}

\begin{figure}
    \centering
    \includegraphics[width=\linewidth]{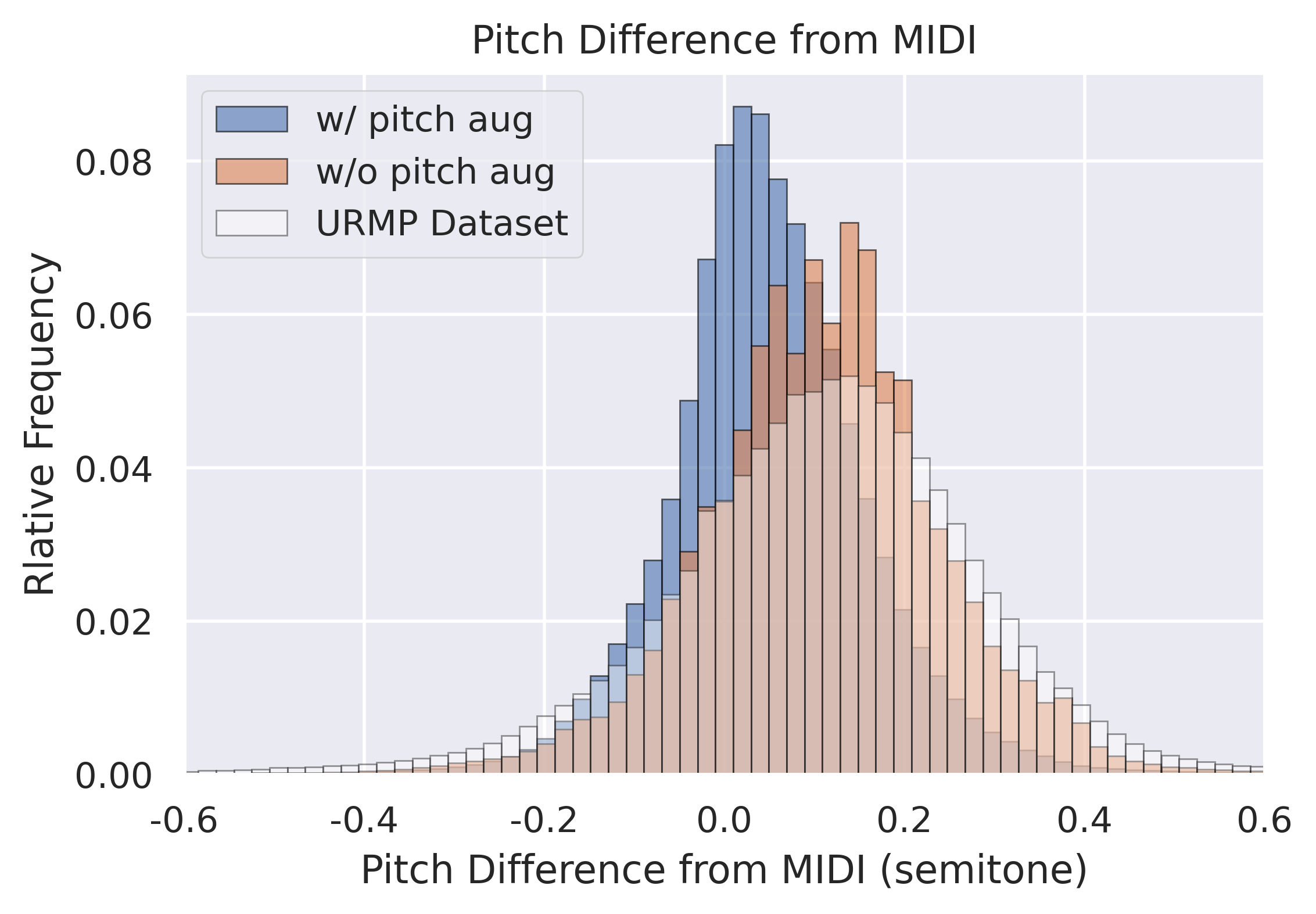}
    \caption{Normalized distributions of $f_0$ values for URMP (white), CocoChorales \textit{without} pitch augmentation (orange) and \textit{with} pitch augmentation (blue), where $f_0$'s are shown as the difference from the closest MIDI pitch (semitones). While a dataset that a generative model is trained on might have systemic biases, our system allows us to correct systemic biases that any of the individual models might introduce, such as producing $f_0$ curves that skew sharp.}
    \label{fig:f0_distributions}
\end{figure}

After the MIDI data is generated, MIDI-DDSP~\cite{wu2022mididdsp} is used to synthesize the MIDI into a realistic-sounding audio performance. Because MIDI-DDSP can only synthesize monophonic audio, each Soprano, Alto, Tenor, and Bass (SATB) part is rendered separately. 

As described in Section~\ref{subsec:midi-ddsp}, MIDI-DDSP offers multiple ways to manipulate the sonic characteristics of its output. 

The first way MIDI-DDSP output can be manipulated is by making edits to the note expression parameters (e.g., note-wise volume, vibrato, etc), and even though it is possible to manually edit these, here we opt to use the expressions that are automatically generated by MIDI-DDSP without manipulation or augmentation.

The second way MIDI-DDSP output can be manipulated is by altering the synthesis parameters, which directly influences the audio output. As mentioned, MIDI-DDSP is trained on the URMP dataset~\cite{li2018creating}. The performances in URMP contains notes that are noticeably sharp, according to twelve-tone equal temperament tuning (12-TET), as shown in the white histogram in Figure~\ref{fig:f0_distributions}. This systematic bias is reflected in the raw $f_0$ curves output by MIDI-DDSP (orange histogram, Figure~\ref{fig:f0_distributions}). To mitigate this, we randomly adjust these intonation deviations for each note by scaling MIDI-DDSP's generated fundamental frequency, $f_0$. We apply a random amount of pitch correction, which helps ensure that the synthesized audio corresponds to the MIDI notes, while allowing a realistic amount of deviation from ``perfect'' 12-TET in the output audio. We hope this could avoid introducing the bias of perfect intonation and makes model trained on CocoChorales robust to intonation errors that are likely to occur in real-life music performances. 

MIDI-DDSP predicts an $f_0$ curve in semitone space as a decimal offset from the in-tune integer pitch value. We randomly transpose the pitch value of each note like so
\begin{equation}
    \hat{f}_0 = f_{0}^{\text{Note}} + \hat{f_{0}^{\Delta}} - \alpha ~
    \bar{f_{0}^{\Delta}}
    \; \text{where} \;
    \alpha \sim \mathcal{U}[0,1],
\end{equation}
and $\hat{f}_0$ is the new fundamental frequency in semitones, $f_{0}^{\text{Note}}$ is the prescribed 12-TET frequency of the note in semitones, $\hat{f_{0}^{\Delta}}$ is the model's predicted $f_0$ offset in semitones, $\bar{f_{0}^{\Delta}}$ is the model's predicted $f_0$ offset in semitones averaged across the note, and $\alpha$ is a random scaling factor. $\alpha=1$, the note is transposed to have a average pitch of the in-tune integer pitch value while $\alpha=0$, the $f_0$ curve is left unchanged from the model prediction.

We synthesize each of the four instrument parts independently and save the audio at 16kHz and 16-bit PCM. We mix each of the four instrument stems following the mixing strategy used by Slakh~\cite{manilow2019cutting}: first, each stem is normalized to have integrated loudness of -13dB, calculated according to the ITU-R BS.1770-4 specification \cite{lkfs2017}. Then, all the stems are summed create an instantaneous mixture. Finally, to prevent clipping, if the summed mixture has a peak loudness larger than -1dB, a uniform gain is applied to the mix (and stems) to ensure that the mix has a peak loudness of exactly -1dB.

\section{Experiments}\label{sec:experiments}

We conduct experiments in multitrack music transcription and source separation to demonstrate the benefits of additional data. Our goal in these experiments is not to propose novel models or architectures for these tasks; instead, it is to demonstrate how the additional data in CocoChorales can be leveraged to improve existing models for the respective tasks.

\subsection{Music Transcription}\label{sec:transcription}

\begin{table}[]
\sisetup{table-format=2.2,round-mode=places,round-precision=2,table-number-alignment = center,detect-weight=true,detect-inline-weight=math}
\centering
\begin{tabular}{lSS}
\toprule 
\textbf{Training Dataset(s)} & {\textbf{On/Off F1}} & {\textbf{Multi-Inst. F1}}  \\ \midrule
URMP & 0.2829 & 0.2224  \\
URMP + CocoChorales &  \B 0.5482 &  \B 0.4417 \\ \bottomrule
\end{tabular}
\caption{Transcription results when training an MT3~\cite{gardner2022mt} model on URMP~\cite{li2018creating} alone versus training one on URMP and CocoChorales, higher is better. URMP is a small dataset (1 hour), so using dataset amplification, as done in here to create CocoChorales, has a large effect on transcription performance.}
\label{tab:transcription-results-urmp}
\end{table}

\begin{table*}
\sisetup{table-format=2.2,round-mode=places,round-precision=2,table-number-alignment = center,detect-weight=true,detect-inline-weight=math}
    \centering
    \hspace*{-\textwidth}\begin{tabular}{@{}lSSSSSS@{}}
        \toprule
        \textbf{Model} & {\textbf{MAESTRO}} &  {\textbf{Cerberus4}} & {\textbf{GuitarSet}} & {\textbf{MusicNet}} & {\textbf{Slakh2100}} & {\textbf{URMP}} \\ \bottomrule
        \multicolumn{7}{c}{ \textbf{\textit{Onset-Offset F1}}}   
        \\ \toprule
        MT3 Datasets & 0.8312 & 0.7987 & 0.7844 & 0.3314 & 0.5702 & 0.6056  \\
        \quad + CocoChorales & 0.8295 & 0.8021 & 0.7889 & 0.3367 & 0.5675 & \B 0.6633 \\ \midrule
        \multicolumn{7}{c}{ \textbf{\textit{Multi-Instrument F1}}}   
        \\ \toprule
        MT3 Datasets         & 0.8312 & 0.7889 & 0.7840 & 0.2980 & 0.5785 & 0.4954 \\
        \quad + CocoChorales &  0.8295 & 0.7496 & 0.7889 & 0.3046 & 0.5744 & \B 0.5598 \\ \bottomrule
    \end{tabular}\hspace*{-\textwidth}
    \caption{Transcription results on the combination of all datasets used by MT3~\cite{gardner2022mt}, higher is better. (Note that we use the MIDI-DDSP \cite{wu2022mididdsp} train-test split for URMP, not the MT3 train-test split for URMP.)}
    \label{tab:overall-results-onset-offset}
\end{table*}

We conducted two sets of music transcription experiments as a demonstration of the effectiveness of our Chamber Ensemble Generator (CEG) and the resulting CocoChorales dataset.

Music transcription, the task of producing a symbolic representation from a raw waveform, is often challenging due to the ``low-resource'' nature of many transcription datasets---that is, high-quality paired data (audio and aligned note annotations) is scarce and expensive to collect. Dataset amplification can provide unlimited data, potentially alleviating this resource limitation. However, whether data from generative models can improve music transcription models is yet unproven.

To explore this, we performed a set of experiments designed to investigate how dataset amplification can improve existing state-of-the-art transcription systems, both in a very low-resource setting (on URMP~\cite{li2018creating}, the dataset that was ``amplified'') and in a combined setting where we train on many existing transcription datasets of various sizes. For all experiments, we used the MT3 transcription model \cite{gardner2022mt}, with no modifications.
For consistency, we used MIDI-DDSP's train/test split of URMP in order to ensure that no inputs used to train the MIDI-DDSP generative model occur in the test set for the transcription model.

Our first study compares the transcription performance of an MT3 model trained only on URMP to an identical model trained on a combined CocoChorales and URMP dataset. This allows us to observe the effectiveness of dataset amplification on the original dataset, URMP. The results of this experiment are shown in Table \ref{tab:transcription-results-urmp}. Our results clearly demonstrate the usefulness of our proposed approach, with the transcription performance on URMP increasing approximately 100\% as measured by both multi-instrument onset-offset F1 and (instrument-agnostic) onset-offset F1 score. For details on the computation of this score, we refer the reader to Raffel, et. al. \cite{raffel2014mir_eval}.

Our second study evaluates whether the addition of the CocoChorales can help improve the model's transcription performance even in the presence of a more diverse collection of datasets. We do this by adding the CocoChorales training dataset to the original dataset combination used to train the MT3 model, and compare an MT3 model trained on the original combination of datasets to an MT3 model trained on the union of those datasets with CocoChorales. This dataset combination, described in detail in the MT3 paper~\cite{gardner2022mt}, contains six standard transcription datasets with a blend of styles, instrumentations, and audio characteristics (i.e. synthesized vs. real).

The results of our second study are shown in Table \ref{tab:overall-results-onset-offset}. These results show that our dataset amplification improves transcription performance on the URMP dataset even over the performance of the state-of-the-art MT3 model trained on the largest available combination of datasets. We note that this improvement from the addition of CocoChorales does not come at a cost to performance on any other datasets, but it also does not improve performance on the datasets besides URMP. To our knowledge, this model also achieves the best published transcription results on the URMP dataset.

\subsection{Source Separation}\label{sec:ssep}

\begin{figure*}[!t]
 \centerline{
 \includegraphics[width=\textwidth]{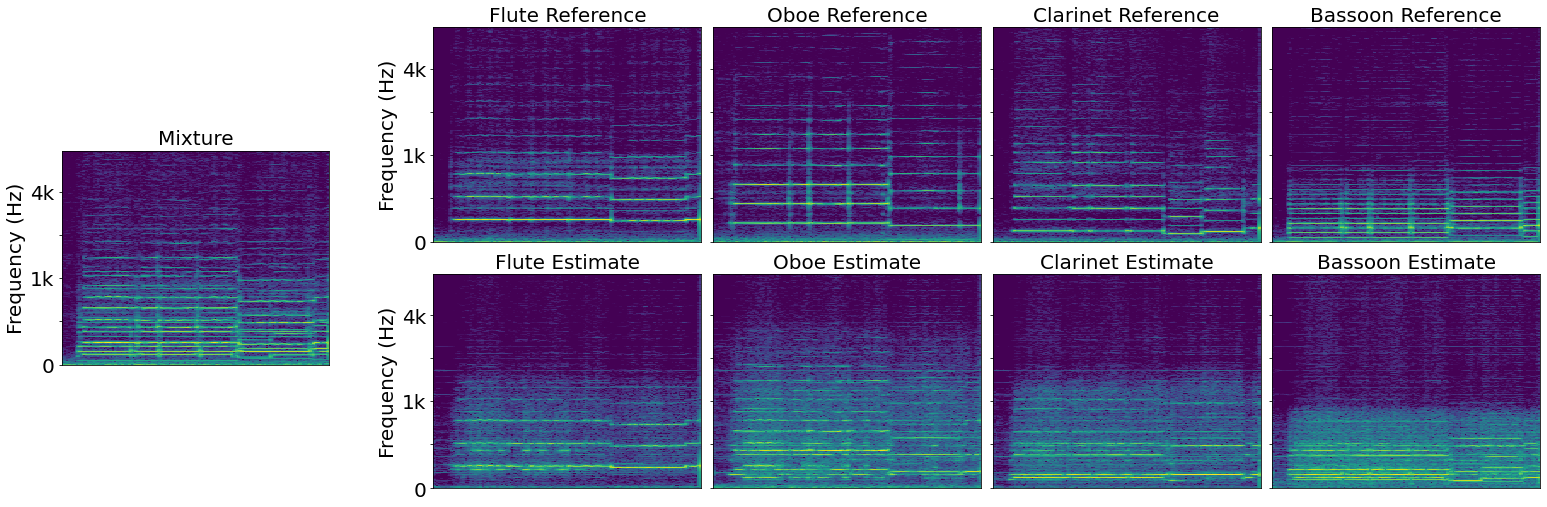}}
 \caption{A Woodwind ensemble from URMP~\cite{li2018creating}, where the mixture (left) is separated by a Demucs v2~\cite{defossez2019music} (bottom row) trained on CocoChorales. The ground truth, reference signals are shown in the top row.} 
 \label{fig:woodwind_separation_example}
\end{figure*}

\begin{table}[]
\sisetup{table-format=2.2,round-mode=places,round-precision=1,table-number-alignment = center,detect-weight=true,detect-inline-weight=math}
\begin{tabular}{lSSSS}
\toprule 
Network   & {Flute}    & {Oboe} & {Clarinet} & {Bassoon} \\ \midrule
Demucs v2~\cite{defossez2019music}  & 18.71  & 17.28 &  21.01  & 20.68 \\
MI+TR~\cite{manilow2020simultaneous} & 12.48  &  6.92 &  10.67  &  6.87 \\ \bottomrule
\end{tabular}
\caption{Mean separation results (dB) for the Woodwind ensemble over the CocoChorales test set using SI-SDR~\cite{le2019sdr}, higher is better.}
\label{tab:woodwind_separation_results}
\end{table}

As a further illustrative example of the usefulness of the Chamber Ensemble Generator (CEG) and resulting CocoChorales dataset, we conduct a set of brief source separation demonstrations trained on the CocoChorales dataset. The majority of music separation systems have focused on separating sources where data is ample, which has historically excluded many instruments (e.g., flutes, oboes, clarinets, or bassoons). Because the CEG enables us to generate a limitless quantity of high-quality ensemble data, we are able to train separation systems on instruments that have been neglected by existing separation systems.

To that end, we train two separation networks to separate instruments from the Woodwind ensemble in CocoChorales. Specifically, we train a Demucs v2~\cite{defossez2019music}, which is a state-of-the-art waveform-to-waveform U-Net, and a Cerberus-style~\cite{manilow2020simultaneous} separation network, which is a set of 4 LSTM layers that create a mask which is applied to the mixture spectrogram. The Cerberus-style network is the same as Cerberus, except we omit the clustering head and only use the Mask Inference and TRanscription heads. We refer to this as MI+TR in Table~\ref{tab:woodwind_separation_results}, and only report its separation performance. For more network details, refer the reader to  \cite{defossez2019music, manilow2020simultaneous}. We train these models for 55k steps and report results over all 5 second segments of the CocoChorales test set in Table~\ref{tab:woodwind_separation_results}, where we report mean SI-SDR~\cite{le2019sdr}.

Furthermore, we also showcase a separation example from URMP~\cite{li2018creating} in Figure~\ref{fig:woodwind_separation_example}. As noted, URMP alone is an extremely small dataset to train a separator on. To train a woodwind ensemble separator, URMP only has two recordings that are woodwind ensembles, totalling less than 4 minutes.
Even if one wanted to just make an oboe separator, URMP has less than 12 \textit{minutes} of oboe data, which would quickly lead to overfitting using modern networks! For this reason, we do not have a URMP-trained baseline for Table~\ref{tab:woodwind_separation_results}.  With CocoChorales, the woodwind ensemble has 360 \textit{hours} of oboe data, which enables us to train separation models that work well on URMP.

\section{Conclusion and Future Applications}

In this paper we introduce the Chamber Ensemble Generator (CEG), which we use to produce the CocoChorales dataset. The CEG is a combination of a generative model for notes (Coconet) and a generative model for audio (MIDI-DDSP) that we set up as a structured hierarchy. In doing this, we can produce an unlimited amount of chamber ensemble mixture data with a rich set of aligned annotation data, with note data, notewise expression data, synthesis parameter data and stem audio data. Using CocoChorales we trained a state-of-the-art transcription system and showed how the data can boost performance on low-resource transcription datasets. We also showed separation results for sources that have been historically underserved by prior separation work.

The experiments we show in Section \ref{sec:experiments} are only two of many possible applications enabled by the Chamber Ensemble Generator and CocoChorales. Because of the rich annotations in the dataset, we are excited by the many other potential applications that this work will enable. Such applications could include, but are not limited to, performance analysis~\cite{lerch2019music,pati2018assessment} (e.g., using the note expressions in the dataset, or deriving new ones from the synthesis parameters), multi-$f_0$ estimation~\cite{bittner2017deep, cuesta2020multiple} (e.g., using the $f_0$'s in this dataset used to synthesize each instrument), or new advances in source separation (e.g., separating multiple instances of similar sounding sources like the string ensemble~\cite{petermann2020deep,schulze2022unsupervised,kawamura2022differentiable}, or separating random ensembles~\cite{lee2019audio,manilow2020hierarchical,chen2021zero}). We look forward to the new directions the MIR community will explore using this data and what new variations on dataset amplification will be explored.








\section{Acknowledgement}
We would like to thank Shuju Han for her help on making the early version of the logo for this work.

\bibliography{ismir2022}

\begin{thebibliography}{10}
\providecommand{\url}[1]{#1}
\csname url@samestyle\endcsname
\providecommand{\newblock}{\relax}
\providecommand{\bibinfo}[2]{#2}
\providecommand{\BIBentrySTDinterwordspacing}{\spaceskip=0pt\relax}
\providecommand{\BIBentryALTinterwordstretchfactor}{4}
\providecommand{\BIBentryALTinterwordspacing}{\spaceskip=\fontdimen2\font plus
\BIBentryALTinterwordstretchfactor\fontdimen3\font minus
  \fontdimen4\font\relax}
\providecommand{\BIBforeignlanguage}[2]{{%
\expandafter\ifx\csname l@#1\endcsname\relax
\typeout{** WARNING: IEEEtran.bst: No hyphenation pattern has been}%
\typeout{** loaded for the language `#1'. Using the pattern for}%
\typeout{** the default language instead.}%
\else
\language=\csname l@#1\endcsname
\fi
#2}}
\providecommand{\BIBdecl}{\relax}
\BIBdecl

\bibitem{kaplan2020scaling}
J.~Kaplan, S.~McCandlish, T.~Henighan, T.~B. Brown, B.~Chess, R.~Child,
  S.~Gray, A.~Radford, J.~Wu, and D.~Amodei, ``Scaling laws for neural language
  models,'' \emph{arXiv preprint arXiv:2001.08361}, 2020.

\bibitem{brown2020language}
T.~Brown, B.~Mann, N.~Ryder, M.~Subbiah, J.~D. Kaplan, P.~Dhariwal,
  A.~Neelakantan, P.~Shyam, G.~Sastry, A.~Askell \emph{et~al.}, ``Language
  models are few-shot learners,'' \emph{Advances in neural information
  processing systems}, vol.~33, pp. 1877--1901, 2020.

\bibitem{henighan2020scaling}
T.~Henighan, J.~Kaplan, M.~Katz, M.~Chen, C.~Hesse, J.~Jackson, H.~Jun, T.~B.
  Brown, P.~Dhariwal, S.~Gray \emph{et~al.}, ``Scaling laws for autoregressive
  generative modeling,'' \emph{arXiv preprint arXiv:2010.14701}, 2020.

\bibitem{bogdanov2019mtg}
D.~Bogdanov, M.~Won, P.~Tovstogan, A.~Porter, and X.~Serra, ``The mtg-jamendo
  dataset for automatic music tagging,'' 2019.

\bibitem{won2019self-attn-tagging}
\BIBentryALTinterwordspacing
M.~Won, S.~Chun, and X.~Serra, ``Toward interpretable music tagging with
  self-attention,'' \emph{CoRR}, vol. abs/1906.04972, 2019. [Online].
  Available: \url{http://arxiv.org/abs/1906.04972}
\BIBentrySTDinterwordspacing

\bibitem{won2020evaluation}
M.~Won, A.~Ferraro, D.~Bogdanov, and X.~Serra, ``Evaluation of cnn-based
  automatic music tagging models,'' \emph{arXiv preprint arXiv:2006.00751},
  2020.

\bibitem{hawthorne2018enabling}
\BIBentryALTinterwordspacing
C.~Hawthorne, A.~Stasyuk, A.~Roberts, I.~Simon, C.-Z.~A. Huang, S.~Dieleman,
  E.~Elsen, J.~Engel, and D.~Eck, ``Enabling factorized piano music modeling
  and generation with the {MAESTRO} dataset,'' in \emph{International
  Conference on Learning Representations}, 2019. [Online]. Available:
  \url{https://openreview.net/forum?id=r1lYRjC9F7}
\BIBentrySTDinterwordspacing

\bibitem{kong2021high}
Q.~Kong, B.~Li, X.~Song, Y.~Wan, and Y.~Wang, ``High-resolution piano
  transcription with pedals by regressing onset and offset times,''
  \emph{IEEE/ACM Transactions on Audio, Speech, and Language Processing},
  vol.~29, pp. 3707--3717, 2021.

\bibitem{hawthorne2021sequence}
C.~Hawthorne, I.~Simon, R.~Swavely, E.~Manilow, and J.~Engel,
  ``Sequence-to-sequence piano transcription with transformers,'' \emph{arXiv
  preprint arXiv:2107.09142}, 2021.

\bibitem{vaswani2017attention}
A.~Vaswani, N.~Shazeer, N.~Parmar, J.~Uszkoreit, L.~Jones, A.~N. Gomez,
  {\L}.~Kaiser, and I.~Polosukhin, ``Attention is all you need,''
  \emph{Advances in neural information processing systems}, vol.~30, 2017.

\bibitem{huang2018music}
C.-Z.~A. Huang, A.~Vaswani, J.~Uszkoreit, N.~Shazeer, I.~Simon, C.~Hawthorne,
  A.~M. Dai, M.~D. Hoffman, M.~Dinculescu, and D.~Eck, ``Music transformer,''
  \emph{arXiv preprint arXiv:1809.04281}, 2018.

\bibitem{payne2019musenet}
C.~Payne, ``Musenet,'' \emph{OpenAI Blog}, 2019.

\bibitem{dhariwal2020jukebox}
P.~Dhariwal, H.~Jun, C.~Payne, J.~W. Kim, A.~Radford, and I.~Sutskever,
  ``Jukebox: A generative model for music,'' \emph{arXiv preprint
  arXiv:2005.00341}, 2020.

\bibitem{castellon2021codified}
R.~Castellon, C.~Donahue, and P.~Liang, ``Codified audio language modeling
  learns useful representations for music information retrieval,'' in
  \emph{Proceedings of 22st International Conference on Music Information
  Retrieval, ISMIR}, 2021.

\bibitem{manilow2022source}
E.~Manilow, P.~O’Reilly, P.~Seetharaman, and B.~Pardo, ``Source separation by
  steering pretrained music models,'' in \emph{ICASSP 2022 - 2022 IEEE
  International Conference on Acoustics, Speech and Signal Processing
  (ICASSP)}, 2022, pp. 126--130.

\bibitem{schluter2015exploring}
J.~Schl{\"u}ter and T.~Grill, ``Exploring data augmentation for improved
  singing voice detection with neural networks.'' in \emph{ISMIR}, 2015, pp.
  121--126.

\bibitem{uhlich2017improving}
S.~Uhlich, M.~Porcu, F.~Giron, M.~Enenkl, T.~Kemp, N.~Takahashi, and
  Y.~Mitsufuji, ``Improving music source separation based on deep neural
  networks through data augmentation and network blending,'' in \emph{2017 IEEE
  International Conference on Acoustics, Speech and Signal Processing
  (ICASSP)}.\hskip 1em plus 0.5em minus 0.4em\relax IEEE, 2017, pp. 261--265.

\bibitem{manilow2019cutting}
E.~Manilow, G.~Wichern, P.~Seetharaman, and J.~Le~Roux, ``Cutting music source
  separation some {Slakh}: A dataset to study the impact of training data
  quality and quantity,'' in \emph{Proc. IEEE Workshop on Applications of
  Signal Processing to Audio and Acoustics (WASPAA)}.\hskip 1em plus 0.5em
  minus 0.4em\relax IEEE, 2019.

\bibitem{defossez2019music}
A.~D{\'e}fossez, N.~Usunier, L.~Bottou, and F.~Bach, ``Music source separation
  in the waveform domain,'' \emph{arXiv preprint arXiv:1911.13254}, 2019.

\bibitem{pretet2019singing}
L.~Pr{\'e}tet, R.~Hennequin, J.~Royo-Letelier, and A.~Vaglio, ``Singing voice
  separation: A study on training data,'' in \emph{ICASSP 2019-2019 ieee
  international conference on acoustics, speech and signal processing
  (icassp)}.\hskip 1em plus 0.5em minus 0.4em\relax IEEE, 2019, pp. 506--510.

\bibitem{manilow2020open}
\BIBentryALTinterwordspacing
E.~Manilow, P.~Seetharman, and J.~Salamon, \emph{Open Source Tools \& Data for
  Music Source Separation}.\hskip 1em plus 0.5em minus 0.4em\relax
  https://source-separation.github.io/tutorial, 2020. [Online]. Available:
  \url{https://source-separation.github.io/tutorial}
\BIBentrySTDinterwordspacing

\bibitem{kong2021decoupling}
Q.~Kong, Y.~Cao, H.~Liu, K.~Choi, and Y.~Wang, ``Decoupling magnitude and phase
  estimation with deep resunet for music source separation,'' \emph{arXiv
  preprint arXiv:2109.05418}, 2021.

\bibitem{song2021catnet}
X.~Song, Q.~Kong, X.~Du, and Y.~Wang, ``Catnet: Music source separation system
  with mix-audio augmentation,'' \emph{arXiv preprint arXiv:2102.09966}, 2021.

\bibitem{spijkervet2021contrastive}
J.~Spijkervet and J.~A. Burgoyne, ``Contrastive learning of musical
  representations,'' \emph{arXiv preprint arXiv:2103.09410}, 2021.

\bibitem{won2021semi}
M.~Won, K.~Choi, and X.~Serra, ``Semi-supervised music tagging transformer,''
  \emph{arXiv preprint arXiv:2111.13457}, 2021.

\bibitem{wang2022towards}
L.~Wang, P.~Luc, Y.~Wu, A.~Recasens, L.~Smaira, A.~Brock, A.~Jaegle, J.-B.
  Alayrac, S.~Dieleman, J.~Carreira \emph{et~al.}, ``Towards learning universal
  audio representations,'' in \emph{ICASSP 2022-2022 IEEE International
  Conference on Acoustics, Speech and Signal Processing (ICASSP)}.\hskip 1em
  plus 0.5em minus 0.4em\relax IEEE, 2022, pp. 4593--4597.

\bibitem{gfeller2020spice}
B.~Gfeller, C.~Frank, D.~Roblek, M.~Sharifi, M.~Tagliasacchi, and
  M.~Velimirovi{\'c}, ``Spice: Self-supervised pitch estimation,''
  \emph{IEEE/ACM Transactions on Audio, Speech, and Language Processing},
  vol.~28, pp. 1118--1128, 2020.

\bibitem{jahanian2022generative}
A.~Jahanian, X.~Puig, Y.~Tian, and P.~Isola, ``Generative models as a data
  source for multiview representation learning,'' in \emph{International
  Conference on Learning Representations}, 2022.

\bibitem{zhang2021datasetgan}
Y.~Zhang, H.~Ling, J.~Gao, K.~Yin, J.-F. Lafleche, A.~Barriuso, A.~Torralba,
  and S.~Fidler, ``Datasetgan: Efficient labeled data factory with minimal
  human effort,'' in \emph{Proceedings of the IEEE/CVF Conference on Computer
  Vision and Pattern Recognition}, 2021, pp. 10\,145--10\,155.

\bibitem{christiano2018supervising}
P.~Christiano, B.~Shlegeris, and D.~Amodei, ``Supervising strong learners by
  amplifying weak experts,'' \emph{arXiv preprint arXiv:1810.08575}, 2018.

\bibitem{axelrod2020sample}
B.~Axelrod, S.~Garg, V.~Sharan, and G.~Valiant, ``Sample amplification:
  Increasing dataset size even when learning is impossible,'' in
  \emph{International Conference on Machine Learning}.\hskip 1em plus 0.5em
  minus 0.4em\relax PMLR, 2020, pp. 442--451.

\bibitem{huang2019counterpoint}
C.-Z.~A. Huang, T.~Cooijmans, A.~Roberts, A.~Courville, and D.~Eck,
  ``Counterpoint by convolution,'' in \emph{Proceedings of 18st International
  Conference on Music Information Retrieval, ISMIR}, 2017.

\bibitem{wu2022mididdsp}
Y.~Wu, E.~Manilow, Y.~Deng, R.~Swavely, K.~Kastner, T.~Cooijmans, A.~Courville,
  C.-Z.~A. Huang, and J.~Engel, ``{MIDI}-{DDSP}: Detailed control of musical
  performance via hierarchical modeling,'' in \emph{International Conference on
  Learning Representations}, 2022.

\bibitem{engel2017neural}
J.~Engel, C.~Resnick, A.~Roberts, S.~Dieleman, M.~Norouzi, D.~Eck, and
  K.~Simonyan, ``Neural audio synthesis of musical notes with wavenet
  autoencoders,'' in \emph{International Conference on Machine Learning}.\hskip
  1em plus 0.5em minus 0.4em\relax PMLR, 2017, pp. 1068--1077.

\bibitem{nikolenko2021synthetic}
S.~I. Nikolenko \emph{et~al.}, \emph{Synthetic data for deep learning}.\hskip
  1em plus 0.5em minus 0.4em\relax Springer, 2021.

\bibitem{zhao2020sim}
W.~Zhao, J.~P. Queralta, and T.~Westerlund, ``Sim-to-real transfer in deep
  reinforcement learning for robotics: a survey,'' in \emph{2020 IEEE Symposium
  Series on Computational Intelligence (SSCI)}.\hskip 1em plus 0.5em minus
  0.4em\relax IEEE, 2020, pp. 737--744.

\bibitem{tremblay2018falling}
J.~Tremblay, T.~To, and S.~Birchfield, ``Falling things: A synthetic dataset
  for 3d object detection and pose estimation,'' in \emph{Proceedings of the
  IEEE Conference on Computer Vision and Pattern Recognition Workshops}, 2018,
  pp. 2038--2041.

\bibitem{wood2021fake}
E.~Wood, T.~Baltru{\v{s}}aitis, C.~Hewitt, S.~Dziadzio, T.~J. Cashman, and
  J.~Shotton, ``Fake it till you make it: Face analysis in the wild using
  synthetic data alone,'' in \emph{Proceedings of the IEEE/CVF International
  Conference on Computer Vision}, 2021, pp. 3681--3691.

\bibitem{karras2019style}
T.~Karras, S.~Laine, and T.~Aila, ``A style-based generator architecture for
  generative adversarial networks,'' in \emph{Proceedings of the IEEE/CVF
  conference on computer vision and pattern recognition}, 2019, pp. 4401--4410.

\bibitem{peracha2021js}
O.~Peracha, ``Js fake chorales: a synthetic dataset of polyphonic music with
  human annotation,'' \emph{arXiv preprint arXiv:2107.10388}, 2021.

\bibitem{liu2020incorporating}
A.~Liu, A.~Fang, G.~Hadjeres, P.~Seetharaman, and B.~Pardo, ``Incorporating
  music knowledge in continual dataset augmentation for music generation,''
  \emph{arXiv preprint arXiv:2006.13331}, 2020.

\bibitem{miron2017generating}
M.~Miron, J.~Janer~Mestres, and E.~G{\'o}mez~Guti{\'e}rrez, ``Generating data
  to train convolutional neural networks for classical music source
  separation,'' in \emph{Lokki T, P{\"a}tynen J, V{\"a}lim{\"a}ki V, editors.
  Proceedings of the 14th Sound and Music Computing Conference; 2017 Jul 5-8;
  Espoo, Finland. Aalto: Aalto University; 2017. p. 227-33.}\hskip 1em plus
  0.5em minus 0.4em\relax Aalto University, 2017.

\bibitem{manilow2020bespoke}
E.~Manilow and B.~Pardo, ``Bespoke neural networks for score-informed source
  separation,'' \emph{arXiv preprint arXiv:2009.13729}, 2020.

\bibitem{chen2022improving}
K.~Chen, H.-W. Dong, Y.~Luo, J.~McAuley, T.~Berg-Kirkpatrick, M.~Puckette, and
  S.~Dubnov, ``Improving choral music separation through expressive synthesized
  data from sampled instruments,'' \emph{arXiv preprint arXiv:2209.02871},
  2022.

\bibitem{cartwright2018increasing}
M.~Cartwright and J.~P. Bello, ``Increasing drum transcription vocabulary using
  data synthesis,'' in \emph{Proc. International Conference on Digital Audio
  Effects (DAFx)}, 2018, pp. 72--79.

\bibitem{boulanger2012modeling}
N.~Boulanger-Lewandowski, Y.~Bengio, and P.~Vincent, ``Modeling temporal
  dependencies in high-dimensional sequences: Application to polyphonic music
  generation and transcription,'' in \emph{International Conference on Machine
  Learning}, 2012.

\bibitem{raffel2016learning}
C.~Raffel, \emph{Learning-based methods for comparing sequences, with
  applications to audio-to-midi alignment and matching}.\hskip 1em plus 0.5em
  minus 0.4em\relax Columbia University, 2016.

\bibitem{ens2021building}
J.~Ens and P.~Pasquier, ``Building the metamidi dataset: Linking symbolic and
  audio musical data,'' in \emph{Proceedings of 22st International Conference
  on Music Information Retrieval, ISMIR}, 2021.

\bibitem{emiya2010maps}
V.~Emiya, N.~Bertin, B.~David, and R.~Badeau, ``Maps-a piano database for
  multipitch estimation and automatic transcription of music,'' 2010.

\bibitem{xi2018guitarset}
Q.~Xi, R.~M. Bittner, J.~Pauwels, X.~Ye, and J.~P. Bello, ``Guitarset: A
  dataset for guitar transcription.'' in \emph{ISMIR}, 2018, pp. 453--460.

\bibitem{foster2021filosax}
D.~Foster, S.~Dixon \emph{et~al.}, ``Filosax: A dataset of annotated jazz
  saxophone recordings,'' in \emph{Proceedings of 22st International Conference
  on Music Information Retrieval, ISMIR}, 2021.

\bibitem{musdb18}
\BIBentryALTinterwordspacing
Z.~Rafii, A.~Liutkus, F.-R. St{\"o}ter, S.~I. Mimilakis, and R.~Bittner, ``The
  {MUSDB18} corpus for music separation,'' Dec. 2017. [Online]. Available:
  \url{https://doi.org/10.5281/zenodo.1117372}
\BIBentrySTDinterwordspacing

\bibitem{thickstun2016learning}
J.~Thickstun, Z.~Harchaoui, and S.~Kakade, ``Learning features of music from
  scratch,'' 2017.

\bibitem{li2018creating}
B.~Li, X.~Liu, K.~Dinesh, Z.~Duan, and G.~Sharma, ``Creating a multitrack
  classical music performance dataset for multimodal music analysis:
  Challenges, insights, and applications,'' \emph{IEEE Transactions on
  Multimedia}, vol.~21, no.~2, pp. 522--535, 2018.

\bibitem{oord2016wavenet}
A.~v.~d. Oord, S.~Dieleman, H.~Zen, K.~Simonyan, O.~Vinyals, A.~Graves,
  N.~Kalchbrenner, A.~Senior, and K.~Kavukcuoglu, ``Wavenet: A generative model
  for raw audio,'' \emph{arXiv preprint arXiv:1609.03499}, 2016.

\bibitem{kalchbrenner2018efficient}
N.~Kalchbrenner, E.~Elsen, K.~Simonyan, S.~Noury, N.~Casagrande, E.~Lockhart,
  F.~Stimberg, A.~Oord, S.~Dieleman, and K.~Kavukcuoglu, ``Efficient neural
  audio synthesis,'' in \emph{International Conference on Machine
  Learning}.\hskip 1em plus 0.5em minus 0.4em\relax PMLR, 2018, pp. 2410--2419.

\bibitem{engel2019gansynth}
J.~Engel, K.~K. Agrawal, S.~Chen, I.~Gulrajani, C.~Donahue, and A.~Roberts,
  ``Gansynth: Adversarial neural audio synthesis,'' \emph{arXiv preprint
  arXiv:1902.08710}, 2019.

\bibitem{goel2022s}
K.~Goel, A.~Gu, C.~Donahue, and C.~R{\'e}, ``It's raw! audio generation with
  state-space models,'' \emph{arXiv preprint arXiv:2202.09729}, 2022.

\bibitem{borsos2022audiolm}
Z.~Borsos, R.~Marinier, D.~Vincent, E.~Kharitonov, O.~Pietquin, M.~Sharifi,
  O.~Teboul, D.~Grangier, M.~Tagliasacchi, and N.~Zeghidour, ``Audiolm: a
  language modeling approach to audio generation,'' \emph{arXiv preprint
  arXiv:2209.03143}, 2022.

\bibitem{ju2021telemelody}
Z.~Ju, P.~Lu, X.~Tan, R.~Wang, C.~Zhang, S.~Wu, K.~Zhang, X.~Li, T.~Qin, and
  T.-Y. Liu, ``Telemelody: Lyric-to-melody generation with a template-based
  two-stage method,'' \emph{arXiv preprint arXiv:2109.09617}, 2021.

\bibitem{chen2020melody}
Y.~Chen and A.~Lerch, ``Melody-conditioned lyrics generation with seqgans,'' in
  \emph{2020 IEEE International Symposium on Multimedia (ISM)}.\hskip 1em plus
  0.5em minus 0.4em\relax IEEE, 2020, pp. 189--196.

\bibitem{liu2022symphony}
J.~Liu, Y.~Dong, Z.~Cheng, X.~Zhang, X.~Li, F.~Yu, and M.~Sun, ``Symphony
  generation with permutation invariant language model,'' \emph{arXiv preprint
  arXiv:2205.05448}, 2022.

\bibitem{wang2019performancenet}
B.~Wang and Y.-H. Yang, ``Performancenet: Score-to-audio music generation with
  multi-band convolutional residual network,'' in \emph{Proceedings of the AAAI
  Conference on Artificial Intelligence}, vol.~33, no.~01, 2019, pp.
  1174--1181.

\bibitem{jonason2020control}
N.~Jonason, B.~Sturm, and C.~Thom{\'e}, ``The control-synthesis approach for
  making expressive and controllable neural music synthesizers,'' in \emph{2020
  AI Music Creativity Conference}, 2020.

\bibitem{castellontowards}
R.~Castellon, C.~Donahue, and P.~Liang, ``Towards realistic midi instrument
  synthesizers,'' in \emph{NeurIPS Workshop on Machine Learning for Creativity
  and Design (2020)}, 2020.

\bibitem{dong2022deep}
H.-W. Dong, C.~Zhou, T.~Berg-Kirkpatrick, and J.~McAuley, ``Deep performer:
  Score-to-audio music performance synthesis,'' in \emph{ICASSP 2022-2022 IEEE
  International Conference on Acoustics, Speech and Signal Processing
  (ICASSP)}.\hskip 1em plus 0.5em minus 0.4em\relax IEEE, 2022, pp. 951--955.

\bibitem{jsb-dataset}
``{JSB-Chorales-dataset},''
  \url{https://github.com/czhuang/JSB-Chorales-dataset}, 2022, [Online;
  accessed 01-May-2022].

\bibitem{coconet-pytorch}
``{Coconet-pytorch},'' \url{https://github.com/lukewys/coconet-pytorch}, 2022,
  [Online; accessed 01-May-2022].

\bibitem{engel2020ddsp}
J.~Engel, L.~Hantrakul, C.~Gu, and A.~Roberts, ``D{D}{S}{P}: Differentiable
  digital signal processing,'' in \emph{International Conference on Learning
  Representations}, 2020.

\bibitem{sogorski2018correlated}
M.~Sogorski, T.~Geisel, and V.~Priesemann, ``Correlated microtiming deviations
  in jazz and rock music,'' \emph{PloS one}, vol.~13, no.~1, p. e0186361, 2018.

\bibitem{naveda2011microtiming}
L.~Naveda, F.~Gouyon, C.~Guedes, and M.~Leman, ``Microtiming patterns and
  interactions with musical properties in samba music,'' \emph{Journal of New
  Music Research}, vol.~40, no.~3, pp. 225--238, 2011.

\bibitem{lkfs2017}
R.~I.-R. BS.1770-4, ``Algorithms to measure audio programme loudness and
  true-peak audio level,'' 2017.

\bibitem{gardner2022mt}
J.~P. Gardner, I.~Simon, E.~Manilow, C.~Hawthorne, and J.~Engel, ``{MT}3:
  Multi-task multitrack music transcription,'' in \emph{International
  Conference on Learning Representations}, 2022.

\bibitem{raffel2014mir_eval}
C.~Raffel, B.~McFee, E.~J. Humphrey, J.~Salamon, O.~Nieto, D.~Liang, D.~P.
  Ellis, and C.~C. Raffel, ``mir\_eval: A transparent implementation of common
  {MIR} metrics,'' in \emph{In Proceedings of the 15th International Society
  for Music Information Retrieval Conference, ISMIR}, 2014.

\bibitem{manilow2020simultaneous}
E.~Manilow, P.~Seetharaman, and B.~Pardo, ``Simultaneous separation and
  transcription of mixtures with multiple polyphonic and percussive
  instruments,'' in \emph{ICASSP 2020-2020 IEEE International Conference on
  Acoustics, Speech and Signal Processing (ICASSP)}.\hskip 1em plus 0.5em minus
  0.4em\relax IEEE, 2020, pp. 771--775.

\bibitem{le2019sdr}
J.~Le~Roux, S.~Wisdom, H.~Erdogan, and J.~R. Hershey, ``Sdr--half-baked or well
  done?'' in \emph{ICASSP 2019-2019 IEEE International Conference on Acoustics,
  Speech and Signal Processing (ICASSP)}.\hskip 1em plus 0.5em minus
  0.4em\relax IEEE, 2019, pp. 626--630.

\bibitem{lerch2019music}
A.~Lerch, C.~Arthur, A.~Pati, and S.~Gururani, ``Music performance analysis: A
  survey,'' \emph{arXiv preprint arXiv:1907.00178}, 2019.

\bibitem{pati2018assessment}
K.~A. Pati, S.~Gururani, and A.~Lerch, ``Assessment of student music
  performances using deep neural networks,'' \emph{Applied Sciences}, vol.~8,
  no.~4, p. 507, 2018.

\bibitem{bittner2017deep}
R.~M. Bittner, B.~McFee, J.~Salamon, P.~Li, and J.~P. Bello, ``Deep salience
  representations for f0 estimation in polyphonic music.''

\bibitem{cuesta2020multiple}
H.~Cuesta, B.~McFee, and E.~G{\'o}mez, ``Multiple f0 estimation in vocal
  ensembles using convolutional neural networks,'' \emph{arXiv preprint
  arXiv:2009.04172}, 2020.

\bibitem{petermann2020deep}
D.~Petermann, P.~Chandna, H.~Cuesta, J.~Bonada, and E.~G{\'o}mez, ``Deep
  learning based source separation applied to choir ensembles,'' \emph{arXiv
  preprint arXiv:2008.07645}, 2020.

\bibitem{schulze2022unsupervised}
K.~Schulze-Forster, C.~S. Doire, G.~Richard, and R.~Badeau, ``Unsupervised
  audio source separation using differentiable parametric source models,''
  \emph{arXiv preprint arXiv:2201.09592}, 2022.

\bibitem{kawamura2022differentiable}
M.~Kawamura, T.~Nakamura, D.~Kitamura, H.~Saruwatari, Y.~Takahashi, and
  K.~Kondo, ``Differentiable digital signal processing mixture model for
  synthesis parameter extraction from mixture of harmonic sounds,'' in
  \emph{ICASSP 2022-2022 IEEE International Conference on Acoustics, Speech and
  Signal Processing (ICASSP)}.\hskip 1em plus 0.5em minus 0.4em\relax IEEE,
  2022, pp. 941--945.

\bibitem{lee2019audio}
J.~H. Lee, H.-S. Choi, and K.~Lee, ``Audio query-based music source
  separation,'' \emph{arXiv preprint arXiv:1908.06593}, 2019.

\bibitem{manilow2020hierarchical}
E.~Manilow, G.~Wichern, and J.~Le~Roux, ``Hierarchical musical instrument
  separation,'' in \emph{International Society for Music Information Retrieval
  (ISMIR) Conference}, 2020, pp. 376--383.

\bibitem{chen2021zero}
K.~Chen, X.~Du, B.~Zhu, Z.~Ma, T.~Berg{-}Kirkpatrick, and S.~Dubnov,
  ``Zero-shot audio source separation through query-based learning from
  weakly-labeled data,'' in \emph{Proceedings of the AAAI Conference on
  Artificial Intelligence}, 2022.

\end{thebibliography}






\end{document}